\documentstyle[12pt]{article}

\def\bc{\begin{center}}
\def\ec{\end{center}}
\def\be{\begin{equation}}
\def\ee{\end{equation}}
\def\bea{\begin{eqnarray}}
\def\eea{\end{eqnarray}}

\newcommand{\lqcd}{\Lambda_{\rm QCD}}

\def\dirule{$\Delta I=1/2$ rule}
\def\oplus{O^{(+)}}
\def\ominus{O^{(-)}}
\def\oplmi{O^{(\pm)}}

\newcommand{\<}{\langle}
\renewcommand{\>}{\rangle}
\def\spose#1{\hbox to 0pt{#1\hss}}
\def\ltapprox{\mathrel{\spose{\lower 3pt\hbox{$\mathchar"218$}}
 \raise 2.0pt\hbox{$\mathchar"13C$}}}
\def\gtapprox{\mathrel{\spose{\lower 3pt\hbox{$\mathchar"218$}}
 \raise 2.0pt\hbox{$\mathchar"13E$}}}
\def\inapprox{\mathrel{\spose{\lower 3pt\hbox{$\mathchar"218$}}
 \raise 2.0pt\hbox{$\mathchar"232$}}}

\begin{document}
\vskip .5cm
\begin{center}{{\LARGE{\bf Lattice approach to the $\Delta I=1/2$ 
rule}}}~\footnote{Talk given at ``Cortona 1998 - Convegno informale di
Fisica Teorica" (Cortona, Italy).}
\end{center}
\vskip 1cm
\centerline{\bf{G.C.~Rossi}}
\vskip 0.3cm
\centerline{Dip. di Fisica, Univ. di Roma ``Tor Vergata''
and INFN, Sezione di Roma II,}
\centerline{Via della Ricerca Scientifica 1, I-00133 Roma, Italy}

\vskip 1cm
\centerline{\bf ABSTRACT}
\begin{quote}

We discuss a number of old and new methods for computing $K\to\pi\pi$
amplitudes on the lattice. They all involve a non-perturbative determination
of matching coefficients. We show how problems related to operator mixing can be
greatly reduced by using point-split hadronic currents.
\end{quote}

\section{Introduction}

One of the major puzzles remaining in hadronic physics is the so-called
``$\Delta I=1/2$ rule" in non-leptonic kaon decays. Decays in which isospin
changes by $\Delta I=1/2$ are greatly enhanced over those with $\Delta
I=3/2$. For instance, one finds experimentally
\begin{equation}
{{\cal A}(K\to\pi\pi[I=0]) \over {\cal A}(K\to\pi\pi[I=2]) } \approx 20 \,.
\label{eq:dirule}
\end{equation}
Although the origin of this large enhancement is not theoretically well
understood, we do know that, in a QCD-based explanation, most of the
enhancement must come from long distance, non-perturbative physics.

Let us briefly discuss the source of the difficulties in
calculating ${\cal A}(K\to\pi\pi)$ in lattice
QCD~\cite{acient}~-~\cite{gavela}.

For scales below $M_{_W}$, but above the charm quark mass, the $\Delta S=1$
part of the effective weak Hamiltonian can be written as 
\bea
{\cal H}_{\rm eff}^{\Delta S=1}
&=&
\lambda_u {G_F \over \sqrt2} 
\left[ C_+(\mu, M_{_W}) \oplus(\mu) + C_-(\mu,M_{_W}) \ominus(\mu) \right]\,,
\label{eq:HW}\\
\oplmi &=& \left[ (\bar s \gamma_\mu^L d)(\bar u \gamma_\mu^L u)
 \pm (\bar s \gamma_\mu^L u)(\bar u \gamma_\mu^L d) \right]
  -  \left[ u \leftrightarrow c \right] \,,
\label{eq:oplmidef}
\eea
where $\gamma_\mu^L=\gamma_\mu (1\!-\!\gamma_5)/2$, 
$\lambda_u=V_{ud}V_{us}^*$, $G_F$ is the Fermi constant and $\mu$ is the
subtraction point.

The operators $\oplmi$ have different transformation properties under isospin.
In particular, $\ominus$ is pure $I=1/2$,
whereas $\oplus$ contains parts having both $I=1/2$ and $I=3/2$.
An explanation of the \dirule\ thus requires
that the $K\to\pi\pi$ matrix element of $C_- \ominus$ 
be substantially enhanced compared  to that of $C_+ \oplus$.

Part of the enhancement is provided by the ratio of Wilson coefficients,
$C_-/C_+$, in their renormalization group evolution from $\mu\sim M_{_W}$ 
down to $\mu \sim 2\;$GeV, where one finds $|C_-/C_+| \approx 2$.
This factor is, however, too small by an order of magnitude to explain the
\dirule. The remainder of the enhancement must come from the matrix elements
of the operators, and these are the quantities that we wish to evaluate
on the lattice. Attempts in this direction date back to the works of 
ref.~\cite{acient}

There are two major problems which arise in the calculation 
of  $\oplmi$ matrix elements in lattice QCD. 
\begin{enumerate}
\item
Decay amplitudes into two or more particles cannot be directly acccessed in
Euclidean space, as a consequence of the Maiani and Testa (MT) no-go
theorem~\cite{MT}, except in kinematical configurations where
final state interactions are absent.  
\item
Operators $\oplmi$ mix with operators of lower dimension with coefficients which
diverge as inverse powers of the lattice spacing. These contributions must
be computed non-perturbatively and subtracted, leading to prohibitively
large statistical errors~\cite{gavela}. 
\end{enumerate}

In this note we briefly revisit the lattice approach for the case of Wilson
fermions (either improved or not)~\footnote{For the use of staggered
fermions see ref.~\cite{toolkit}.}, recalling old methods and reviewing 
new proposals~\cite{noinew}.

\section{Old and new approaches}

There is a long list of methods proposed in the literature to deal with the
problem of computing the hadronic matrix elements that enter in the
non-leptonic weak decay amplitude on the lattice. They are all aimed at
bypassing the two major difficulties illustrated in the previous section. 
Here we briefly highlight the merits and the drawbacks of the most promising
among them.

1) In ref.~\cite{direct} it was suggested to work
with $m_s=m_d$ and calculate the Euclidean amplitude 
\be 
{\cal A}_{m_s=m_d}^{(\pm)} = \langle \pi(\vec p_1\!=\!0) \pi(\vec
p_2\!=\!0)   | O^{(\pm)}(\mu) | K(\vec p_K\!=\!0) \rangle\bigg|_{m_s=m_d} 
\label{eq:b+s}
\ee
with all three particles at rest. Using GIM and CPS-symmetry (CPS=
CP$\times$ $s\leftrightarrow d$ interchange \cite{politzer}), it can be shown
that setting $m_s=m_d$ causes all mixings to vanish, removing the
need for subtractions. Working with the two pions at rest solves the
problem of final state interactions, as they vanish at threshold. The method
requires, however, a large extrapolation from the unphysical point $m_s=m_d$
to the physical one with the use of chiral perturbation theory~\cite{GL}. 

2) An alternative method~\cite{noinew} consists in working with a
non-perturbatively $O(a)$ improved fermion action (for which there are no
errors of $O(a)$ in the spectrum~\cite{SWH} and on-shell
matrix elements of improved currents obey the continuum chiral Ward
identities up to $O(a^2)$~\cite{luscher}). Choosing quark masses such that $m_K=
2m_\pi$, one measures the $K\to\pi\pi$ amplitudes with all particles at
rest. Since with this choice of masses the momentum transfer vanishes
($\Delta p=0$), one can prove that the matrix element of the dangerous,
divergent, subtraction is now of $O(a)$ rather than of $O(1)$ and vanishes in
the continuum limit. Furthermore pions are at rest and the no-go MT theorem
does not apply. Besides the need of a (perhaps less severe) chiral extrapolation, a
problem here is the difficulty of tuning with a sufficiently good accuracy
the quark masses to have the condition $m_K= 2m_\pi$ satisfied.

3) The old proposal of ref.~\cite{MMRT} makes use of the $K\to\pi$
matrix elements of (the positive parity part of) the weak Hamiltonian and
relies on chiral perturbation theory, in the form of Soft Pion Theorems
(SPT's), to connect them to $K\to\pi\pi$ amplitudes. Since only
single-particle states are involved, there are no problems with the
MT theorem. The disadvantage of the method is that the operator
mixing problem is much more severe than for the negative parity part of
${\cal H}_{\rm eff}^{W}$ (which enters in the  $K\to\pi\pi$ amplitudes), thus
making an accurate evaluation of the matrix elements of the renormalized
operator very difficult.

The relation between $K\to\pi\pi$ and $K\to\pi$ amplitudes is easily found
exploiting SPT's. At leading order in chiral perturbation theory the physical
amplitude takes, in fact, the form (for $\Delta p=0$) 
\be
\<\pi^+\pi^-\vert  O^{(\pm)}(\mu)\vert K^0\> =
i\,\gamma^{(\pm)}\, {m^2_K - m^2_\pi\over f_\pi} 
\,. \label{eq:SPT3}
\ee
As the coefficients $\gamma^{(\pm)}$ appear also in the expression for
the $K\to\pi$ matrix element 
\begin{equation}
\<\pi^+(p)\vert  O^{(\pm)}(\mu) \vert K^+(q)\> =
-\delta^{(\pm)}\, {m^2_K\over f_\pi^2} + \gamma^{(\pm)}\,p\cdot q
\,.\label{eq:SPT2}
\end{equation}
by studying this matrix element on the lattice as a function of $p\cdot q$, 
one can, in principle, determine $\gamma^{(\pm)}$, from which we obtain the
$K\to\pi\pi$ amplitudes.

In order to construct the finite renormalized lattice operators $O^{(\pm)}$ 
it was suggested in reference~\cite{MMRT} to use perturbation theory to
determine finite mixing coefficients and subtract non-perturbatively
operators with divergent mixing coefficients. In the case of the parity
conserving part of $O^{(\pm)}$ that contribute to eq.~(\ref{eq:SPT2}), this 
can be elegantly done by adjusting the only divergent coefficient until the 
momentum independent part of the $K\to\pi$ matrix element ($\delta^{(\pm)}$ in
eq.~(\ref{eq:SPT2})) vanishes.

4) A method which, in principle, avoids all the difficulties caused by
mixing with lower dimension operators and which automatically gives the
effective weak Hamiltonian with the correct normalization is based on the
idea of studying at short distances ($a\ll |x|  \ll 
\Lambda_{QCD}^{-1}$) the $x$-behaviour of the OPE expansion of two hadronic
currents on the lattice.

The method does not use chiral perturbation theory, and thus applies
equally well to the $\Delta S=1$, $\Delta C=1$ and $\Delta B=1$ parts of
the weak Hamiltonian. In addition, it allows one to construct an improved weak
Hamiltonian (i.e.~one having errors of $O(a^2)$), if the improved version of
the weak hadronic currents~\cite{luscher} are used.

The method is speculative in the sense that it is likely to require
more computational power than is presently available, although we expect it to
become practical with the advent of Teraflop machines.

We recall that the standard construction of the non-leptonic weak Hamiltonian
begins with the formula 
\begin{equation}
{\cal H}_{\rm eff}^{W}=g^2_{_W}\int d^4x\,D_{_W} (x;M_{_W}) 
T\Big(J_{\rho L}(x) J^\dagger_{\rho L}(0)\Big) \, ,
\label{eq:HEFF}
\end{equation}
where
\begin{equation}
D_{_W}(x;M_{_W})=\int d^4p\,\frac{\mbox{e}^{ipx}}{p^2+M_{_W}^2} 
\label{WPROP}
\end{equation}
is the longitudinal part of the $W$-boson propagator and $J_{\rho L}$ is the
(left-handed) hadronic weak current. One then introduces the operator
product expansion (OPE) in the r.h.s of eq.~(\ref{eq:HEFF}), which is
justified by the observation that the dominant contribution to the integral
comes from distances $|x| \ll M_{_W}^{-1}$. For physical amplitudes, one
obtains in this way 
\begin{equation}
\langle h|{\cal H}_{\rm eff}^{W}|h'\rangle =
\frac{G_{F}}{\sqrt{2}} \sum_i C_i(\mu,M_{_W}) M_{_W}^{6-d_i} 
\langle h|{O}^{(i)}(\mu)|h'\rangle\ ,
\label{eq:HEFFOPE}
\end{equation}
where $d_i$ is the dimension of the operator ${O}^{(i)}(\mu)$,
and the functions $C_i(\mu,M_{_W})$ result from the integration
of the Wilson expansion coefficients, $c_i(x;\mu)$ (defined in 
eq.~(\ref{eq:ME}) below), with the $W$-propagator,
\begin{equation}
C_i(\mu,M_{_W}) M_{_W}^{6-d_i} 
= \int d^4x\,D_{_W} (x;M_{_W}) c_i(x;\mu)\ .
\label{eq:WILCOEF}
\end{equation}
The ${O}^{(i)}(\mu)$ are quark and/or gluon operators
renormalized at the subtraction point $\mu$. The functions $C_i(\mu,M_{_W})$
are evaluated in perturbation theory and their running with $\mu$ is dictated
by the renormalization group equation which follows from the
$\mu$-independence of the l.h.s. of eq.~(\ref{eq:HEFFOPE}). 

The sum in the expansion~(\ref{eq:HEFFOPE}) is over operators of
increasing dimension. As the operator dimension of $\cal {H}_{\rm eff}^{W}$ 
is 6, we will have to consider in the following only operators with dimensions 
$d_i \le 6$, since the contribution from operators with $d_i>6$ is suppressed 
by powers of $1/M_{_W}$.

All the intricacies of operator mixing in the definition of the finite
and renormalized operators, ${O}^{(i)}(\mu)$, come about because the
integrals in~(\ref{eq:HEFF}) and~(\ref{eq:WILCOEF}) are extended down
to the region of extremely small $x$. The complicated mixing of the
${O}^{(i)}(\mu)$'s in terms of bare operators arises from contact
terms when the separation of the two currents goes to zero, i.e.~when
$|x|$ is of the order of $a$ (the problem is particularly bad here, because
chiral symmetry is broken by the lattice regularization). This
observation suggests that a simple way to circumvent these difficulties is
to directly determine the matrix elements of renormalized operators by
enforcing the OPE on the lattice for distances $|x|$ much larger than the
lattice spacing $a$, but much smaller than $\Lambda_{QCD}^{-1}$, i.e.~in a
region where perturbation theory can be used to determine the expected form of
the OPE. 

We imagine proceeding in the following way. If $J_{\rho L}$ is the 
appropriately renormalized (and possibly improved) finite lattice current
operator, we measure in Monte Carlo simulations the hadronic matrix element 
$< h\vert T(J_{\rho L}(x) J^\dagger_{\rho L}(0))\vert h'\>$, as a function 
of $x$ in the region  $a\ll |x|  \ll  \Lambda_{QCD}^{-1}$. The 
numbers $\< h\vert {O}^{(i)}(\mu)\vert h'\>$ entering in eq.~(\ref{eq:HEFFOPE}) 
are extracted by fitting the $x$-behaviour of 
$\< h\vert T(J_{\rho L}(x) J^\dagger_{\rho L}(0))\vert h'\>$ to the OPE formula  
\begin{equation}
\< h\vert T\Big (J_{\rho L}(x) J^\dagger_{\rho L}(0)\Big)\vert h' \> = \sum_i
c_i(x;\mu) \< h\vert {O}^{(i)}(\mu)\vert h' \> 
\,,\label{eq:ME}
\end{equation}
where the Wilson coefficients $c_i(x;\mu)$ are determined by
continuum perturbation theory using any renormalization scheme we
like. The scale $\mu$ should be chosen so that $1/\mu$ too lies in the
range $a\ll  1/\mu \ll  \Lambda_{QCD}^{-1}$. Since 
we only consider operators of dimension 6 or lower, the lattice $T$-product
differs from the right-hand side of eq.~(\ref{eq:ME}) by terms of
$O(|x|^2\lqcd^2)$, which is then an estimate of the size of the systematic
errors intrinsic in this procedure. As a last step we insert the numbers $\<
h\vert {O}^{(i)}(\mu)\vert h'\>$, determined in this way,
in~(\ref{eq:HEFFOPE}),  finally obtaining an explicit expression for the
matrix elements of  ${\cal H}_{\rm eff}^{W}$.

The procedure illustrated above requires the existence of a window, $a\ll |x| 
\ll  \Lambda_{QCD}^{-1}$, in which the distance between the two currents is
sufficiently small that perturbation theory can be used, but large enough that
lattice artifacts, which are suppressed by powers of $a/x$, are tiny. For such
a window to exist we need to have an adequately small lattice spacing.
At the same time the physical volume of the lattice must be sufficiently
large to allow the formation of hadrons.

A few remarks may be useful at this point.
\begin{itemize}
\item
The method determines directly the ``physical'' matrix elements of the
operators appearing in the OPE of the  two currents, i.e. the matrix elements
of the finite, renormalized operators ${O}^{(i)}(\mu)$, without any reference
to the magnitude of the $W$-mass. This means that it will not be necessary to
probe distances of $O(1/M_{_W})$ with lattice calculations.
\item
Since it is the continuum OPE which determines the operators appearing in the
lattice OPE~(\ref{eq:HEFFOPE}), these are restricted by the continuum
symmetries. This is because, for $|x|\gg a$, the lattice OPE matches that of
the continuum with discretization errors suppressed by powers of $a/x$.
\item 
The $\mu$-dependence of the matrix elements of the operators
${O}^{(i)}(\mu)$ is trivially induced by that of the (perturbative)
Wilson coefficients, $c_i(x;\mu)$.  It compensates the related
$\mu$-dependence of the functions $C_i(\mu,M_{_W})$ in such a way that
the l.h.s of eq.~(\ref{eq:HEFFOPE}) is independent of the choice of the
subtraction point.
\item 
Unlike the methods discussed before, this approach automatically yields
hadronic amplitudes that are properly  normalized (in the renormalization
scheme in which the  Wilson coefficients appearing in eq.~(\ref{eq:ME}) are
computed). 
\end{itemize}

As for the applicability of this method, we remark that, fortunately, in
the case at hand there are no operators of dimension lower than 6. If lower 
dimension operators were present they would dominate at short distances,
since their Wilson coefficients would diverge as powers of $1/x$ (up to
logarithmic corrections). In this situation it would be virtually impossible
to pick out the matrix elements of the interesting dimension 6 operators.

Operators of dimension 6 have, instead, Wilson coefficients which vary
logarithmically with $x$. At leading order their expression is 
\be
c_i(x;\mu) \propto 
\left(\alpha_s(1/x) \over \alpha_s(\mu)\right)^{\gamma_0^{(i)}\over 2 \beta_0}
= 1 + \frac{\alpha_s}{4\pi}\gamma_0^{(i)} \log(x \mu) + \dots
\,,
\label{eq:formofci}
\ee
where $\gamma_0^{(i)}$ is the one-loop anomalous dimension of the operator
$O^{(i)}$, and $\beta_0$ is coefficient of the one-loop term in the
$\beta$-function.

In the case of the $\Delta S=1$ part of ${\cal H}_{\rm eff}^{W}$, the
operators which can appear in~(\ref{eq:HEFFOPE}) are $\oplmi$ 
(eq.~(\ref{eq:oplmidef})) and in addition 
\be O' = (m_c^2-m_u^2) \,
\bar s (\overrightarrow{D_\mu} - \overleftarrow{D_\mu}) \gamma_\mu^L d
\,.
\ee
The GIM mechanism causes $O'$ to vanish when $m_c=m_u$, while chiral
symmetry requires both the quarks to be left-handed and that the GIM
factor be quadratic in the quark masses. Since $O'$ has dimension 6, its
coefficient function depends only logarithmically on $x$.\par

The anomalous dimensions of the three relevant operators ($\oplmi$ and $O'$) 
are
\be
\gamma^{(+)}_0 = 4, \qquad 
\gamma^{(-)}_0 = -8, \qquad
\gamma '_0 = 16\ .
\label{eq:ANOMALDIM}
\ee

Actually, the contribution of $O'$ to the r.h.s. of eq.~(\ref{eq:ME})
can be determined separately, since its matrix element does  
not require any subtraction, and needs not be fitted. The anomalous
dimensions of the operators $O^{(\pm)}$ are well separated from one another,
so it might be possible to determine the amplitudes $\< h\vert \oplmi
(\mu)\vert h' \>$ and obtain the physical matrix elements of ${\cal H}_{\rm
eff}^{\Delta S=1}$.

\section{Conclusions}
\label{sec:concs}

In this note we have reviewed a number of new approaches aimed at studying the
\dirule\ on the lattice, using Wilson-like fermions (similar methods could also
be used for staggered fermions).

We have reevaluated the methods of refs.~\cite{direct} and~\cite{MMRT}
involving $K\to\pi\pi$ and $K\to\pi$ amplitudes, respectively. The
last approach is likely to be more difficult than the first one, because of
the large number of mixing coefficients which have to be determined
non-perturbatively. It may however give complementary information to the
results obtained with the $K\to\pi\pi$ methods and provide a check of the
accuracy of chiral relations.

The recently proposed approach~\cite{noinew} based on the study of the short
distance behaviour of the OPE of two hadronic weak currents on the lattice is
theoretically very appealing. The advent of Teraflop Supercomputers may 
render it a viable method to directly extract physical amplitudes from Monte
Carlo data. An interesting feasibility study in this direction has been
undertaken~\cite{MONT}. It consists in the investigation of the small 
$x$ behaviour of the OPE of two currents in the two-dimensional O(3) 
$\sigma$-model. A preliminary analysis of Monte Carlo data indicates 
that the measured small $x$ behaviour of the one-particle matrix elements of 
the product of two currents matches the logarithmic behaviour expected from 
perturbative calculations, thus allowing the (non-perturbative) evaluation 
of the matrix elements of the operators appearing in the OPE.

\end{document}